\documentstyle[12pt]{article}
\def\fr{\frac}
\newcommand{\bq}{\begin{equation}}
\newcommand{\eq}{\end{equation}}
\begin{document}
\bigskip\bigskip
{\raggedleft {\makebox[2.7cm][1]{\large\it ASITP-92-45}}\\}
\bigskip
{\raggedleft {\makebox[5.0cm][1]{\large\it SISSA-ISAS 147/92/EP}}\\}
\bigskip
{\raggedleft {\makebox[2.7cm][1]{\large\it August 1992}}\\}
\bigskip\bigskip\bigskip
\centerline{\Large\bf
Intrinsic Vertex Regularization and Renormalization\\}
\bigskip
\centerline{\Large\bf in $\phi^4$ Field Theory\\}

\begin{center}

\bigskip\bigskip\bigskip\bigskip
{\large Zhong-Hua Wang}\\

\bigskip

Institute of Theoretical Physics, Academia Sinica, Beijing 100080,
China.\\

 and\\

 SISSA, Strada Costiera 11, 34014 Trieste, Italy.\\
\bigskip\bigskip

{\large Han-Ying Guo}\\

\bigskip

Institute of theoretical physics, Academia Sinica, Beijing 100080, China.\\

\bigskip\bigskip\bigskip\bigskip

\end{center}
\vskip 2pt

\medskip
\centerline{\it ABSTRACT}
\bigskip\bigskip
\begin{center}
\begin{minipage}{120mm}
{\it Based upon the intrinsic relation between the divergent lower point
functions and the convergent higher point ones in the renormalizable quantum
field theories, we propose a new method for regularization and renormalization
in QFT. As an example, we renormalize the $\phi^{4}$ theory at the one loop
order by means of this method.}\\

\end{minipage}
\end{center}
\bigskip\bigskip\bigskip\bigskip
\vskip 2pt
\eject

\bigskip
\bigskip
\bigskip

\bigskip

 In the last decades, various regularization and renormalization schemes
have been developed in the quantum field theory [1]. However, the  problem is
still one of the most important issues under investigation in the modern QFT.

 In this letter we propose a new method for regularization and renormalization
in QFT and take the $\phi^4$ field theory at one loop order as
an example to show the spirit of the method. The main idea is based upon some
simple observations in the renormalizable QFT. First, there
are always, to a definite loop order, a limited number of lower point
correlation functions which are divergent while
the higher point functions are convergent and well defined if the
number of points is bigger enough. Secondly, there exist certain intrinsic
relations between the lower point functions and the higher point ones.
Namely, the lower point functions can be reached as  limiting cases of
the number of the points, or the vertices, in  the higher point convergent
function which are well defined. This is in fact the crucial property we want
to use in our  method.

  The main points of our method are as follows. In order to
regularize a one particle irreducible (1PI) $n$ point divergent function, we
first calculate the $n+2q$  point functions
$\Gamma^{(n+2q)}(p_1,~\cdots,~p_n,~\underbrace{0,~\cdots,~0}_{2q})$ where
the momenta of $2q$ external lines are set to zero.
When $q$ is large enough, $\Gamma^{(n+2q)}$ is convergent and well defined. It
can also be expressed in terms of $p_1,~\cdots,~p_n$
and $q$.
Then we introduce $\Gamma^{(n)}(p_1,~\cdots,~p_n;~q;~\mu)\equiv\mu^{2q}
\Gamma^{(n+2q)}(p_1,~\cdots,~p_n,~\underbrace{0,~\cdots,~0}_{2q})$ where $\mu$
is a constant with the dimension of mass in order that
$\Gamma^{(n)}(p_1,~\cdots,~p_n;~q;~\mu)$ has the same dimension
as the 1PI $n$ point function. After calculating
$\Gamma^{(n)}(p_1,~\cdots,~p_n;~q;~\mu)$, we make an analytic continuation
of $q$ from the integer to the complex number and it is called in our scheme
the regularized $n$ point function. The original 1PI n-point function is
recovered when $q\to 0$:
\begin{equation}
\Gamma^{(n)}(p_1,~\cdots,~p_n;\mu)=\lim_{q\to 0}
\Gamma^{(n)}(p_1,~\cdots,~p_n;~q;~\mu)
\end{equation}
We call this kind of regularization in our scheme the intrinsic vertex
regularization.
The renormalization of the theory in this method is the same as in the usual
approach. Namely,
we subtract the divergent part of the vertices of the theory at each loop
order
by adding the relevant counterterms to the original action. The new action is
the renormalized one. The renormalized $n$ point
functions are then calculated from the renormalized action. When $q\to 0$, we
get a finite result for all the correlation functions.

Let us now concentrate on the divergent diagrams in $\phi^4$ theory at the one
loop order. We first show that in the $\phi^4$ theory  the divergent parts of
such  regularized $n$ point functions at the one loop order behaves as
$q^{-1}$ and the convergent parts tend to finite quantities when $q\to 0$.

 The action of the theory is
\begin{equation}
S[\phi]=\int d^4 x~(\fr{1}{2}\partial_{\mu}\phi\partial^{\mu}\phi +\fr{1}{2}
m^2 \phi^2 +\fr{\lambda}{4!}\phi^4)
\end{equation}
where $\lambda$ is the coupling constant.
The Feynman rules for the $\phi^4$ theory are well known:\\
\vskip 3mm
{}~~~Propagator:~~~~~~~~~~~~~~~~~~~~~~~~~~~~~~~~$=i(p^2-m^2)^{-1}$\\
\vskip 5mm
{}~~~Vertex:~~~~~~~~~~~~~~~~~~~~~~~~~~~~~~~ $=-i\lambda$\\
\vskip 3mm
The superficial degree of divergence for a proper vertex with $n$ external
lines (1PI n-point correlation function) is
\begin{equation}
\delta=4-n
\end{equation}
Hence the superficially divergent proper vertices at the one loop order are
depicted as the diagrams $(a),~ (b),~ (c)$ and $(d)$  in Fig. 1.

In the momentum space, the amplitude of these vertices are
\begin{equation}
\begin{array}{ll}
(a)=\fr{1}{2}\int \fr{d^4 l}{(2\pi)^4} ~~\fr{\lambda}{l^2-m^2}\\[5mm]
(b)=\fr{1}{2}\int \fr{d^4 l}{(2\pi)^4}~~
\fr{\lambda^2}{(l^2-m^2)((p_1+p_2+l)^2-m^2)}\\[4mm]
(c)=(b)(p_1+p_2\to p_1+p_3),~~~~~~(d)=(b)(p_1+p_2\to p_1+p_4)
\end{array}
\end{equation}
It is easy to see from (4) that $(a)$ is quadratically divergent and $(b),~(c)$
and $(d)$ are logarithmically divergent. We can also see from $(3)$ that
the diagrams with $n>4$ external lines at the one loop order are
superficially convergent.

\bigskip

{\large\it Regularization and evaluation of Feynman integrals}:

 In order to make the integral expressions for $(a),~(b),~(c)$ and
$(d)$  to be meaningful, we have to regularize them. The regularization
procedure is as follows:

First, we attach to the loop, say in the diag.$(a)$, $q$
vertices or $2q$ extra
external lines with the momentum of each external line being zero in order
that the diag. $(a)$ turns to the diag.$(a')$ ( Fig.2) .
Then we introduce a dimensional constant $\mu$
with the dimension of mass in order that the dimension of $(a)$
equals to the dimension of $\mu^{2q}(a')$ and
when $q= 0$, $\mu^{2q}(a')=(a)$.

The amplitude  $\mu^{2q}(a')$ can then be expressed as
\begin{equation}
\mu^{2q}(a')=\mu^{2q}\fr{1}{2}\int \fr{d^4 l}{(2\pi)^4} ~~
\fr{\lambda^{q+1}}{(l^2-m^2)^{q+1}}
\end{equation}
For $q$ large enough, $\mu^{2q}(a')$ is
convergent and is well defined. Furthermore,
the amplitude $\mu^{2q}(a')$ can be easily integrated and expressed
in terms of the $Gamma$
functions of $q$:
\begin{equation}
\mu^{2q}(a')=\mu^{2q}\fr{i}{2}\fr{\lambda^{q+1}}{(4\pi)^{2}}\fr{\Gamma(q-1)}
{\Gamma(q+1)(-m^2)^{q-1}}.
\end{equation}
This expression does also make sense when we take the continuation of $q$
from the integer  to the complex number. And when $q\to 0$, we have
\begin{equation}
\lim_{q\to 0} \mu^{2q}(a')=\fr{i}{2}\fr{\lambda m^2}{(4\pi)^2}[\fr{1}{q}
+1+\ln (-\fr{\lambda\mu^2}{m^2})+o(1)].
\end{equation}
where $o(1)$ represents those terms that are regular in $q$.

Obviously, this  means that $\mu^{2q}(a')$
can be defined as the regularization of $(a)$ and the procedure of the
regularization is completed.

 Similarly, we attach to one of the internal lines of diag. $(b)$  $q$
vertices or $2q$ external lines with zero momentum and
the diag. $(b)$ turns to the diag. $(b')$ ( Fig. 3).
The dimension of $\mu^{2q}(b')$
 is the same
as that  of $(b)$ and when $q=0$, $\mu^{2q}(b')=(b)$.

The amplitude of $\mu^{2q}(b')$ is
\begin{equation}
\mu^{2q}(b')=\fr{\mu^{2q}}{2}\int \fr{d^4 l}{(2\pi)^4}~~
\fr{\lambda^{2+q}}{(l^2-m^2)^{q+1}((p_1+p_2+l)^2-m^2)}
\end{equation}
 For $q$ large enough, $\mu^{2q}(b')$ is
convergent and is well defined.

The integration of $\mu^{2q}(b')$ can be performed by using the Feynman
parametrization
\begin{equation}
\begin{array}{ll}
\mu^{2q}(b')=\lambda^2\fr{(\mu^2\lambda)^q}{2}
\int_0^1 d\alpha~(q+1)\alpha^q\int \fr{d^4 l}{(2\pi)^4}~~
\fr{1}{(l^2-m^2+(1-\alpha)[(p_1+p_2)^2+2l\cdot(p_1+p_2)])^{q+2}}\\[4mm]
=i\lambda^2\fr{(\mu^2\lambda)^q}{2}\int_0^1 d\alpha~(q+1)\alpha^q
\fr{\Gamma(q)}{(4\pi)^2\Gamma(q+2)}\Bigl(-m^2+\alpha(1-\alpha)(p_1+p_2)^2
\Bigr)^{-q}\\[4mm]
=\fr{i\lambda^2}{(4\pi)^2}\fr{(\mu^2\lambda)^q}{2q}\int_0^1 d\alpha~\alpha^q
\Bigl(-m^2+\alpha(1-\alpha)(p_1+p_2)^2\Bigr)^{-q}
\end{array}
\end{equation}
{}From this expression, we make the analytic continuation of $q$ from the
integer to the complex number and we regard $\mu^{2q}(b')$ as the
regularized form of (b). When $q\to 0$, we find
\begin{equation}
\begin{array}{ll}
\lim_{q\to 0}\mu^{2q}(b')\\[4mm]
=\fr{i\lambda^2}{(4\pi)^2}\Bigl(\fr{1}{2q}+\fr{1}{2}+\fr{1}{2}
\ln (\fr{-\lambda \mu^2}{m^2})-\fr{1}{2}\sqrt{1-\fr{4m^2}{(p_1+p_2)^2}}
\ln \fr{\sqrt{1-\fr{4m^2}{(p_1+p_2)^2}}+1}{\sqrt{1-\fr{4m^2}{(p_1+p_2)^2}}-1}+
o(1)\Bigr).
\end{array}
\end{equation}

We can also get the regularized expressions of the diag. $(c)$ and the
diag.$(d)$. It is easy to see that
they have the same behavior as the diag.$(b)$ except that $p_1+p_2$ substituted
by $p_1+p_3$ and $p_1+p_4$ respectively.

 Hence we get the divergent proper vertices  at the one loop order:
\begin{equation}
\begin{array}{ll}
\Gamma^{(2)}(p)=i(p^2-m^2)+\fr{i}{2}\fr{\lambda m^2}{(4\pi)^2}[\fr{1}{q}
+1+\ln (-\fr{\lambda\mu^2}{m^2})+o(q)] \\[4mm]
\Gamma^{(4)}(p_1,~p_2,~p_3,~p_4)\\[4mm]
=-i\lambda
+\fr{i\lambda^2}{(4\pi)^2}\Bigl(\fr{3}{2q}+\fr{3}{2}+\fr{3}{2}
\ln (\fr{-\lambda \mu^2}{m^2})-A(p_1+p_2)-A(p_1+p_3)-A(p_1+p_4)\Bigr)+
o(q)
\end{array}
\end{equation}
where
$$
A(p_1+p_2)=
\fr{1}{2}\sqrt{1-\fr{4m^2}{(p_1+p_2)^2}}
\ln \fr{\sqrt{1-\fr{4m^2}{(p_1+p_2)^2}}+1}{\sqrt{1-\fr{4m^2}{(p_1+p_2)^2}}-1}.
$$
We find that infiniteness of these divergent proper  vertices resulted in poles
in $q$. Furthermore, the finite part of these proper vertices are arbitrary
depending in our scheme on the mass parameter $\mu$.

\bigskip
{\large\it Renormalization at the one loop order}:

In order to eliminate these divergencies in $q$ at the one loop order, we may
add some counterterms
to the original action. These counterterms, in general, can be written as
\begin{equation}
\fr{1}{2}(Z_{\phi}-1)\partial_{\mu}\phi\partial^{\mu}\phi
+\fr{1}{2}\delta m^2\phi^2+\fr{1}{4!}\lambda(Z_{\lambda}-1) \phi^4
\end{equation}
where $Z_{\phi}$, $\delta m^2$ and $Z_{\lambda}$ are some constants
depending on $q$, $m$, $\lambda$ and $\mu$.

Hence the renormalized action $S_r(\phi)$ is:
\begin{equation}
\begin{array}{ll}
S_r[\phi]=\int d^4 x~\Bigl(\fr{1}{2}\partial_{\mu}\phi\partial^{\mu}\phi
 +\fr{1}{2}
m^2 \phi^2 +\fr{\lambda}{4!}\phi^4\\[4mm]
{}~~~~~~~~~~~~~~~+\fr{1}{2}(Z_{\phi}-1)\partial_{\mu}\phi\partial^{\mu}\phi
+\fr{1}{2}\delta m^2\phi^2+\fr{1}{4!}\lambda(Z_{\lambda}-1) \phi^4\Bigr)
\end{array}
\end{equation}
If we choose $Z_{\phi}$, $\delta m^2$ and $Z_{g}$ to be
\begin{equation}
\begin{array}{ll}
Z_{\phi}=1,\\[4mm]
\delta m^2=\fr{\lambda m^2}{2 (4\pi)^2}\fr{1}{q}+o(1),\\[4mm]
Z_{\lambda}=1+\fr{3\lambda}{2 (4\pi)^2}\fr{1}{q}+o(1),
\end{array}
\end{equation}
where $o(1)$ represents those terms  that are regular in $q$ and can be
determined by the renormalization conditions, we find that all the
correlation functions of $\phi^4$ theory have a finite small $q$ limit
at the one loop order.

If we redefine
$$
\phi_0=\sqrt Z_{\phi}\phi,~~~m^2_0=Z_{\phi}^{-1}(m^2+\delta m^2),~~~
\lambda_0=Z_{\phi}^{-2}Z_{\lambda}\lambda,
$$
the renormalized action (13) can be rewritten as
\begin{equation}
S_r(\phi_0)=\int d^4x~\Bigl(\fr{1}{2}\partial_{\mu}\phi_0\partial^{\mu}\phi_0
+\fr{1}{2}m^2_0 \phi_0^2 +\fr{\lambda_0}{4!}\phi_0^4\Bigr),
\end{equation}
where $\phi_0$, $m_0$ and $\lambda_0$ are the so-called bare quantities.
The physical quantities $\phi$, $m$ and $\lambda$ can be
expressed in terms of these bare ones[2].

 The bare 1PI $n$ point function
$\Gamma_0^{(n)}(p_1,~\cdots,~p_n;~\lambda_0,~m_0,~q)$,
relate with the renormalized function
$\Gamma^{(n)}_r(p_1,~\cdots,~p_n;~\lambda,~m,~\mu,~q)$ through
\begin{equation}
\Gamma_0^{(n)}(p_1,~\cdots,~p_n;~\lambda_0,~m_0,~q)=Z_{\phi}^{-\fr{n}{2}}
\Gamma^{(n)}_r(p_1,~\cdots,~p_n;~\lambda,~m,~\mu,~q)
\end{equation}
We note that the left hand side of eq. (16)  is
independent of $\mu$. Based upon this observation, it is easy to get the
renormalization group equation:
\begin{equation}
[\mu\fr{\partial}{\partial \mu}+\beta (\lambda,\fr{m}{\mu},q)\fr{\partial}
{\partial \lambda}
+\gamma_m(\lambda,\fr{m}{\mu},q)\fr{\partial}{\partial m}-\fr{n}{2}
\gamma_d(\lambda,\fr{m}{\mu},q)]
\Gamma^{(n)}_r(p_1,\cdots,p_n;\lambda,m,\mu,q)=0
\end{equation}
where
$$
\beta (\lambda,\fr{m}{\mu},q)=\mu\fr{\partial\lambda}{\partial \mu},~~
\gamma_m(\lambda,\fr{m}{\mu},q)=\fr{\mu}{2}\fr{\partial \ln m^2}
{\partial \mu},~~
\gamma_d(\lambda,\fr{m}{\mu},q)=\fr{\mu}{2}\fr{\partial \ln Z_{\phi}}
{\partial \mu}.
$$
Further discussions on the renormalization group equation are the same as
that in the dimensional regularization method.

The new method of regularization and renormalization proposed here is very
simple and useful. It works not only for the $\phi^4$ theory but also for
the QED, chiral fermionic theory, QCD and other field theory models.
The relevant results will be presented elsewhere [3].

\bigskip\bigskip

  The authors are grateful to  L. Bonora and C. S. Xiong for discussions.
HYG would like to thank SISSA and Professor L. Bonora for the warm hospitality
extended to him during his visiting to SISSA. ZHW is obliged to
Professors D. Amati, L. Bonora, R. Iengo for their warm hospitality during his
working in SISSA sponsored by the World Laboratory.

\bigskip
{\raggedright \bf References\\}
\begin{description}
\bigskip
\item[1] See, for example, C. Itzykson and J.-B. Zuber, ``Quantum Field
Theory",  McGraw-Hill Inc. and references therein.
\item[2] P. Ramond, ``Field Theory: A Modern Primer", 2nd edition,
Addison-Wesley Publishing Company, Inc.
\item[3] Zhong-Hua Wang  and H. Y. Guo, in preparation.

\end{description}

\end{document}